\PassOptionsToPackage{breaklinks=true}{hyperref}
\PassOptionsToPackage{capitalize}{cleveref}
\documentclass[]{fairmeta}

\usepackage{enumitem}
\usepackage{tabularx}
\usepackage{array}
\usepackage{tcolorbox}
\tcbuselibrary{breakable}

\usepackage[utf8]{inputenc}
\usepackage[T1]{fontenc}
\usepackage{microtype}
\usepackage{graphicx}
\usepackage{subcaption}
\usepackage{wrapfig}
\usepackage{booktabs}
\usepackage{xurl}
\usepackage{amsfonts}
\usepackage{nicefrac}
\usepackage{xcolor}

\usepackage{tikz}
\usetikzlibrary{arrows.meta,positioning,fit}

\usepackage{amsmath}
\usepackage{amssymb}
\usepackage{mathtools}
\usepackage{amsthm}

\title{Superintelligent Retrieval Agent: The Next Frontier of Agentic Retrieval}

\author[1,2,*]{Zeyu Yang}
\author[1]{Xu Han}
\author[1]{Qi Ma}
\author[1]{Jason Chen}
\author[1,2,*]{Anshumali Shrivastava}

\affiliation[1]{Meta Superintelligence Labs}
\affiliation[2]{Rice University}

\contribution[*]{Work done at Meta}

\abstract{
Retrieval-augmented agents are increasingly the interface to large
organizational and public knowledge bases, yet most still treat retrieval as a
black box: they issue exploratory queries, inspect returned snippets, and
iteratively reformulate until useful evidence emerges. This resembles how a
newcomer searches an unfamiliar database rather than how an expert navigates it
with strong priors about terminology, constraints, and likely evidence, leading
to unnecessary retrieval rounds, increased latency, and poor recall.

We introduce \textit{Superintelligent Retrieval Agent} (SIRA), which defines
\emph{superintelligence} in retrieval as the ability to compress multi-round
exploratory search into a single corpus-discriminative retrieval action. SIRA
does not merely ask what terms are relevant to the query; it asks which terms
are likely to separate the desired evidence from corpus-level confusers. On the
corpus side, an LLM enriches each document offline with missing search
vocabulary; on the query side, it predicts evidence vocabulary omitted by the
query; and corpus statistics are used as tool calls to filter proposed terms
that are absent, overly common, or unlikely to create retrieval margin. The
final retrieval step is a single weighted BM25 call combining the original query
with the validated expansion.

Across ten BEIR benchmarks, SIRA achieves the strongest average retrieval
performance in our comparison, outperforming dense retrievers, learned sparse
retrievers, and LLM-based search-agent baselines while using no relevance labels
or retriever fine-tuning. On downstream question answering, SIRA's
retrieval-only answer coverage exceeds recent RL-trained agentic QA systems on
NQ and HotpotQA. Finally, we introduce \textbf{BrowseComp-Wikipedia}, a
hard-search benchmark of 232 BrowseComp-derived queries grounded in a
25,587,229-document English Wikipedia index. Even without index-time LLM
document enrichment, using only grounded Wikipedia categories as corpus-visible
structure, SIRA outperforms multi-round Perplexity agents at every retrieval
budget, reaching 9.70\% Recall@1, 15.27\% Recall@10, and 36.14\% Recall@100.
These results show that one well-formed, corpus-grounded lexical retrieval
action can outperform substantially more expensive multi-round search while
remaining interpretable, training-free, and efficient.
}

\date{\today}
\correspondence{Zeyu Yang at \email{zy45@meta.com} and Anshumali Shrivastava \email{anshumali@meta.com}}

\metadata[Code]{Available at \url{https://github.com/facebookresearch/sira}}
\metadata[BrowseComp-Wikipedia Dataset]{Available at \url{https://github.com/facebookresearch/sira}}

\begin{document}

\maketitle

\section{Introduction}
\label{sec:introduction}

Information retrieval (IR) has evolved from lexical matching, exemplified by BM25~\citep{robertson2009probabilistic}, to neural retrieval dominated by dense embeddings~\citep{karpukhin2020dense}.
Embedding-based retrievers perform well when trained with abundant in-domain supervision: large-scale relevance labels and interaction logs that calibrate the model to a platform's user population~\citep{bajaj2016ms,joachims2007evaluating,chapelle2009dynamic}.
This regime has fueled modern \emph{retrieval-augmented generation} (RAG) systems that ground LLM outputs in external corpora~\citep{lewis2020retrieval}.

The user interface for information access is changing rapidly: search is increasingly \emph{answer-forward} and \emph{conversational}, with LLMs mediating multi-turn information seeking~\citep{mo2025survey}.
A key consequence is that the classic training signal for supervised ranking, clickthrough, becomes sparse, delayed, and biased: users often terminate sessions without clicking, or accept an on-page summary as the final answer.
A large-scale browsing analysis by Pew Research Center finds that when Google presents an AI-generated summary, users click standard result links substantially less often~\citep{pew2025aisummaries}.
Click-based supervision is therefore becoming unreliable at scale, precisely as query behavior is shifting.

\paragraph{Compositional queries need controllable retrieval.}
At the same time, \emph{query distributions} are moving away from short keyword strings toward longer, compositional requests that combine constraints, exclusions, and multi-step intent.
This shift is a hallmark of conversational search.
Pure similarity search is an awkward fit for this regime: dense retrieval exposes only a black-box nearest-neighbor operator and provides weak handles for enforcing structure (e.g., must-include/must-not-include constraints, attribute filters, or explicit decomposition).
Neural sparse methods such as SPLADE~\citep{formal2021splade} partially restore lexical controllability while preserving learning-based ranking, but they are still used as fixed retrievers inside pipelines rather than as controllable components of an agent policy.

Classical lexical retrieval, exemplified by BM25, possesses underappreciated strengths that become decisive when paired with LLM reasoning.
BM25 is \emph{transparent}: an agent can boost keywords, enforce constraints, and decompose queries with predictable effects on retrieval outcomes.
It naturally rewards \emph{rare, discriminative terms} via IDF weighting, so domain-specific jargon that would be diluted in a dense embedding becomes a powerful retrieval signal.
It is \emph{auditable}: one can trace exactly which keywords matched and why, while avoiding the latency and memory costs of dense indices.
The missing ingredient has been a mechanism to surface the right rare terms and constraints; LLMs, with their vast parametric knowledge, are uniquely positioned to fill this role.

\paragraph{LLM reasoning meets retrieval.}
The limitations of dense retrieval are not merely engineering inconveniences; they expose a deeper mismatch between single-vector representations and compositional information needs.
Recent theoretical and empirical work shows that fixed-dimensional embeddings can realize only a limited family of relevance patterns~\citep{weller2025theoretical}, that static embeddings are information bottlenecks whenever relevance requires cross-attention-style interaction~\citep{anshu2025attentionembeddinglimits}, and that vector databases impose substantial cost, latency, and objective-mismatch burdens in production~\citep{thirdai2023vectorlimits} 

In parallel, LLM reasoning frameworks such as Chain-of-Thought, Tree-of-Thoughts, and Graph-of-Thoughts demonstrate that LLMs can plan and explore structured intermediate states~\citep{wei2022chain,yao2023tree,besta2024graph}.
Tool-using agents extend this capability to external actions, including search~\citep{yao2022react}, and recent reinforcement-learning approaches train LLMs to interleave reasoning with multi-turn web search~\citep{jin2025search}.
In most agentic search systems, however, retrieval itself remains an opaque tool: the agent can rewrite queries and judge snippets, but cannot directly manipulate retrieval primitives such as keyword weighting, constraint selection, or decomposition tied to index-time signals.

This exposes why current search agents remain brittle despite strong reasoning capabilities. Agentic systems such as ReAct, IRCoT, and recent RL-trained search agents improve performance by interleaving reasoning with repeated retrieval calls \citep{yao2022react,trivedi2023interleaving,jin2025search}. However, this success is partly obtained through a \emph{retrieval-context advantage}: after each search, the agent absorbs returned snippets, discovered entities, surface vocabulary, and near misses into the LLM context, then uses this accumulated evidence to formulate later queries. In other words, the agent compensates for weak retrieval control by learning the corpus through interaction. This strategy is expensive and noisy, and it relies increasingly on long-context LLMs to retain and use many intermediate passages---a regime known to be unreliable when relevant evidence is buried in long contexts \citep{liu2024lost}. Thus the failure mode is not simply that LLMs cannot reason; it is that the search interface gives them too little direct control, forcing them into hit-and-miss exploration.

\paragraph{The known hardness of retrieval that LLM agents are clueless about.} Classical IR theory clarifies the missing ingredient. Retrieval is not just a question of whether a query is semantically related to the desired document; it is a comparative ranking problem in which the gold evidence must outrank many non-gold \emph{confusers}. The probability ranking principle and learning-to-rank methods formalize this as ordering relevant documents above non-relevant alternatives \citep{robertson1977probability,joachims2007evaluating,chapelle2009dynamic}, while BM25 operationalizes corpus contrast through document frequency and IDF \citep{robertson2009probabilistic}. A query can therefore be plausible in isolation yet fail because its terms also match many distractors, or because its most expert-sounding terms are absent, too common, or weakly discriminative in the target index. Dense single-vector retrieval adds a further bottleneck: recent theory shows that fixed-dimensional embedding retrievers cannot realize all top-$k$ relevance patterns and can fail even on simple realistic constraint structures \citep{weller2025theoretical}. This points to the core gap SIRA addresses: an LLM may know what relevant evidence should look like, but it needs index-visible statistics and explicit retrieval controls to make that expectation discriminative against the confusing articles that corpus may be flooded with.

SIRA formalizes the setting of one-shot, controllable BM25 retrieval across broad IR and QA benchmarks, building on recent evidence that LLM agents can generate discriminative keyword, grep, and ripgrep queries approaching RAG-level QA performance~\citep{subramanian2025keyword,wang2026greprag}.
Existing results, however, are either practitioner-level, code-centric, or multi-turn and context-accumulating~\citep{karpathy2026llmwiki,cognition2025swegrep}; SIRA targets the stronger single-query regime.

A particularly important stress test is hard web-style search. BrowseComp was
introduced to measure whether browsing agents can locate hard-to-find,
short-answer facts on the web~\citep{wei2025browsecomp}. Its questions are
valuable because they are difficult to find but easy to verify, but the original
benchmark relies on live web search, where the corpus, ranking function, and
snippets are uncontrolled. BrowseComp-Plus improves reproducibility by
constructing a fixed corpus of 100,195 documents for 830 verified queries
~\citep{chen2025browsecomp}. However, a curated $10^5$-document corpus is
still much smaller than the search spaces faced by real agents, and it weakens
the core disambiguation problem: ranking the right page above millions of
plausible confusers. We therefore introduce BrowseComp-Wikipedia, a
232-query benchmark derived from BrowseComp and grounded entirely in a
25,587,229-document English Wikipedia index. This setting preserves the
hard-search character of BrowseComp while making retrieval controlled,
public-corpus grounded, and substantially larger scale.

\begin{figure}[t]
\centering
\includegraphics[width=\linewidth]{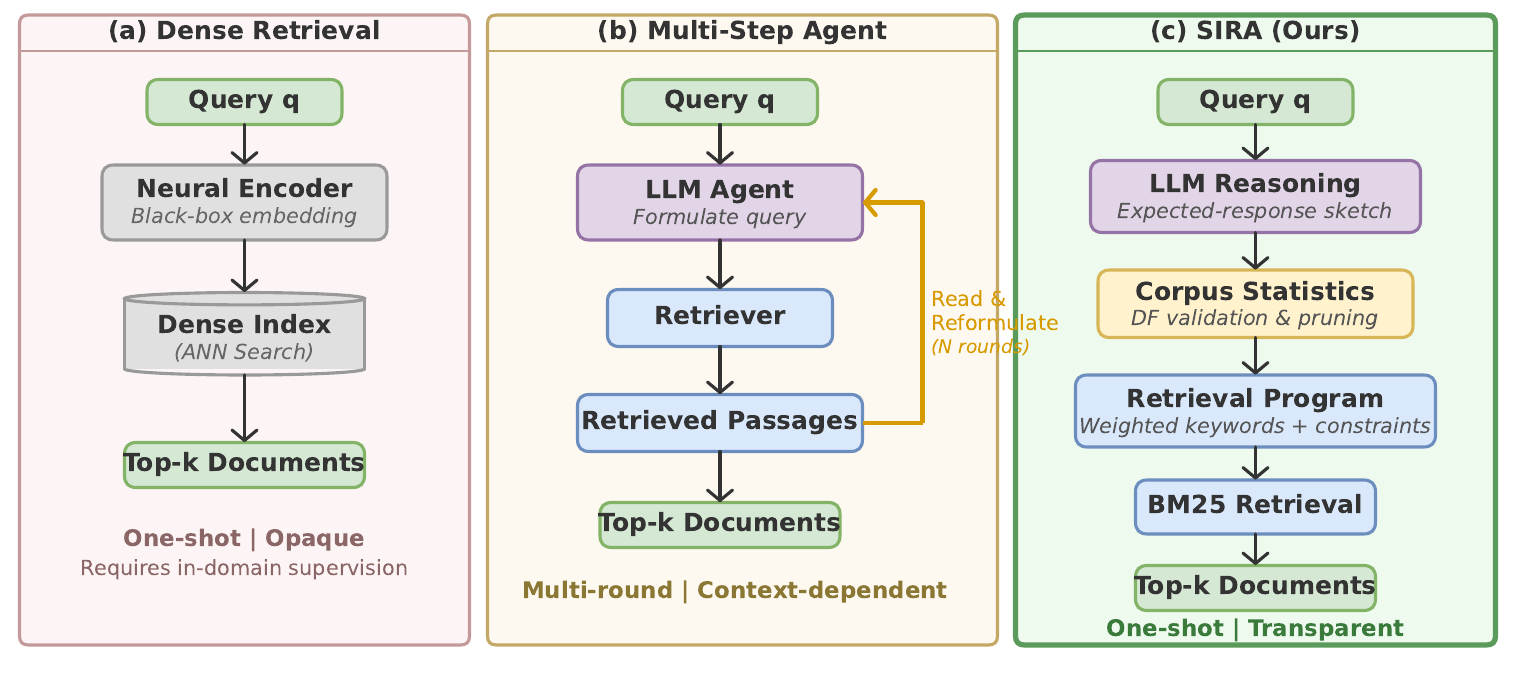}
\caption{Three retrieval paradigms compared.
\textbf{(a)}~Dense retrieval encodes queries and documents into a shared embedding space and performs nearest-neighbor search; the process is one-shot but opaque and requires in-domain supervision.
\textbf{(b)}~Multi-step agent retrieval uses an LLM to iteratively formulate queries, read retrieved passages, and reformulate over $N$ rounds; later queries benefit from accumulated retrieval context.
\textbf{(c)}~SIRA produces an expert-level retrieval action in a single shot: the LLM generates an expected-response sketch, validates proposed terms against corpus statistics, and compiles a controlled BM25 query with weighted keywords and constraints, all without reading any retrieved passages.}
\label{fig:paradigm-comparison}
\end{figure}

\paragraph{Goal.}
The central limitation of today's LLM-driven retrieval agents is that retrieval remains a \emph{black-box environment} they explore through repeated interaction.
Existing agents issue a query, inspect returned snippets, and reformulate using accumulated evidence, a \emph{retrieval-context advantage} where later searches are conditioned on information exposed by earlier ones.
This resembles a domain newcomer learning an unfamiliar database through exploration, rather than an expert who anticipates relevant terminology and evidence patterns before reading retrieved passages, as illustrated in \cref{fig:paradigm-comparison}.

We define \emph{superintelligence in retrieval} as the ability to replace this multi-round process with a single expert-level retrieval action:
(i)~form a domain-informed expectation of what relevant evidence looks like,
(ii)~ground that expectation using lightweight index-aware signals (document frequency),
(iii)~compile the result into explicit retrieval controls, and
(iv)~execute retrieval efficiently and transparently.

\subsection{Our Proposal: SIRA}
We propose the \emph{Superintelligent Retrieval Agent} (SIRA), a retrieval-centric agent that searches like a domain expert.
Modern LLMs encode substantial parametric knowledge such as terminology, entities, relations, but lack a mechanism to convert this knowledge into precise retrieval actions over a target corpus.
SIRA provides that mechanism through a scalable two-stage framework.

First, the LLM produces an \emph{expected-response sketch}: a compact hypothesis of the concepts, entities, and discriminative terms likely to appear in relevant evidence.
This sketch acts as a retrieval prior, not as evidence.
Before issuing the final BM25 query, SIRA consults lightweight corpus-statistics tools (document frequencies) to validate and prune proposed terms without returning answer passages, avoiding the retrieval-context advantage enjoyed by multi-round agents.

Second, conditioned on the sketch and index statistics, SIRA compiles a \emph{retrieval program}: a single controlled BM25 query with weighted keywords, optional exclusions, and structured composition.
We evaluate SIRA in this strict one-shot setting: one LLM reasoning step, optional index-statistic checks, and one BM25 call.
Multiple queries could improve performance, but the one-query regime isolates the central question: how far can retrieval go when the agent must formulate the right lexical action without reading retrieved snippets?

BM25's reliance on exact matching becomes a strength when the LLM supplies the right vocabulary and verifies its discriminative value through corpus statistics.

We evaluate SIRA in three increasingly demanding settings. First, across ten
BEIR-style retrieval benchmarks~\citep{thakur2021beir}, SIRA outperforms strong
dense retrievers, learned sparse retrievers, and LLM-based retrieval-agent
baselines on average Recall@10 and NDCG@10. Second, on downstream QA, SIRA's
retrieval-only answer coverage exceeds recent RL-trained agentic QA systems on
NQ and HotpotQA, even though those systems use trained search policies and
answer generators. Third, on BrowseComp-Wikipedia, SIRA directly challenges the
prevailing multi-round search-agent paradigm: a single corpus-grounded BM25
query, constructed before reading retrieved pages, outperforms Perplexity-based
agents that can browse iteratively for up to 100 turns. These results suggest
that the bottleneck in retrieval-augmented agents is not the sophistication of
the reader or the number of search iterations, but the agent's ability to
construct an expert-level, corpus-discriminative retrieval action.

\section{Background and Organization}
\label{sec:background}

\paragraph{BM25 and sparse lexical retrieval.}
BM25~\citep{robertson2009probabilistic} is the dominant lexical ranking function in modern search systems.
It scores a document~$d$ against a query~$q = (q_1, \ldots, q_n)$ by summing per-term contributions:
\begin{equation}
\label{eq:bm25}
\text{BM25}(q, d) = \sum_{i=1}^{n} \underbrace{\log\!\left(1 + \frac{N - n(q_i) + 0.5}{n(q_i) + 0.5}\right)}_{\text{IDF}(q_i)} \;\cdot\; \frac{f(q_i, d)}{f(q_i, d) + k_1 \!\left(1 - b + b \cdot \tfrac{|d|}{\text{avgdl}}\right)},
\end{equation}
where $f(q_i, d)$ is the frequency of term~$q_i$ in~$d$, $|d|$ is the document length in tokens, $\text{avgdl}$ is the average document length across the corpus, $N$ is the corpus size, and $n(q_i)$ is the number of documents containing~$q_i$.

The formula decomposes into two interpretable factors.
We use the Lucene variant, which applies $\log(1+x)$ to the classical Robertson--Sp\"arck Jones ratio, guaranteeing non-negative IDF for all terms.
The IDF term down-weights common words and up-weights rare, discriminative terms: a query term appearing in most documents contributes near-zero IDF, while one appearing in only a handful of documents receives a large weight.
The TF saturation term ensures that repeating a word within a document yields diminishing returns, controlled by~$k_1$.
The parameter~$b$ governs length normalization: at $b{=}1$ the score fully normalizes for document length, while at $b{=}0$ document length is ignored.

BM25 is implemented over an inverted index, which maps each vocabulary term to the documents and term frequencies in which it appears.

\paragraph{Index-visible signals.}
BM25 makes corpus contrast observable through the inverted index: before retrieval, an agent can check whether a candidate term appears in the corpus, how many documents contain it, and how much IDF weight it can contribute. These signals do not reveal answer passages, but they expose whether LLM-proposed vocabulary is absent, too common, or likely to create retrieval margin. SIRA uses this information to validate, prune, and weight expansion terms so the final query is not merely plausible in isolation, but discriminative within the target corpus.

\section{SIRA: Superintelligent Retrieval Agent}
\label{sec:sira}

\subsection{Overview}
\label{sec:overview}

Most retrieval-augmented agents interact with a search engine through a loop: issue a query, inspect results, reformulate, and repeat until useful evidence emerges.
SIRA replaces this multi-round loop with a \emph{one-shot} pipeline that bridges the vocabulary gap between queries and documents from \emph{both sides} simultaneously.
The full system requires no training, no relevance labels, and no supervised query--document pairs; it operates with a frozen LLM and corpus statistics alone.

\begin{figure}[t]
\centering
\includegraphics[width=\textwidth]{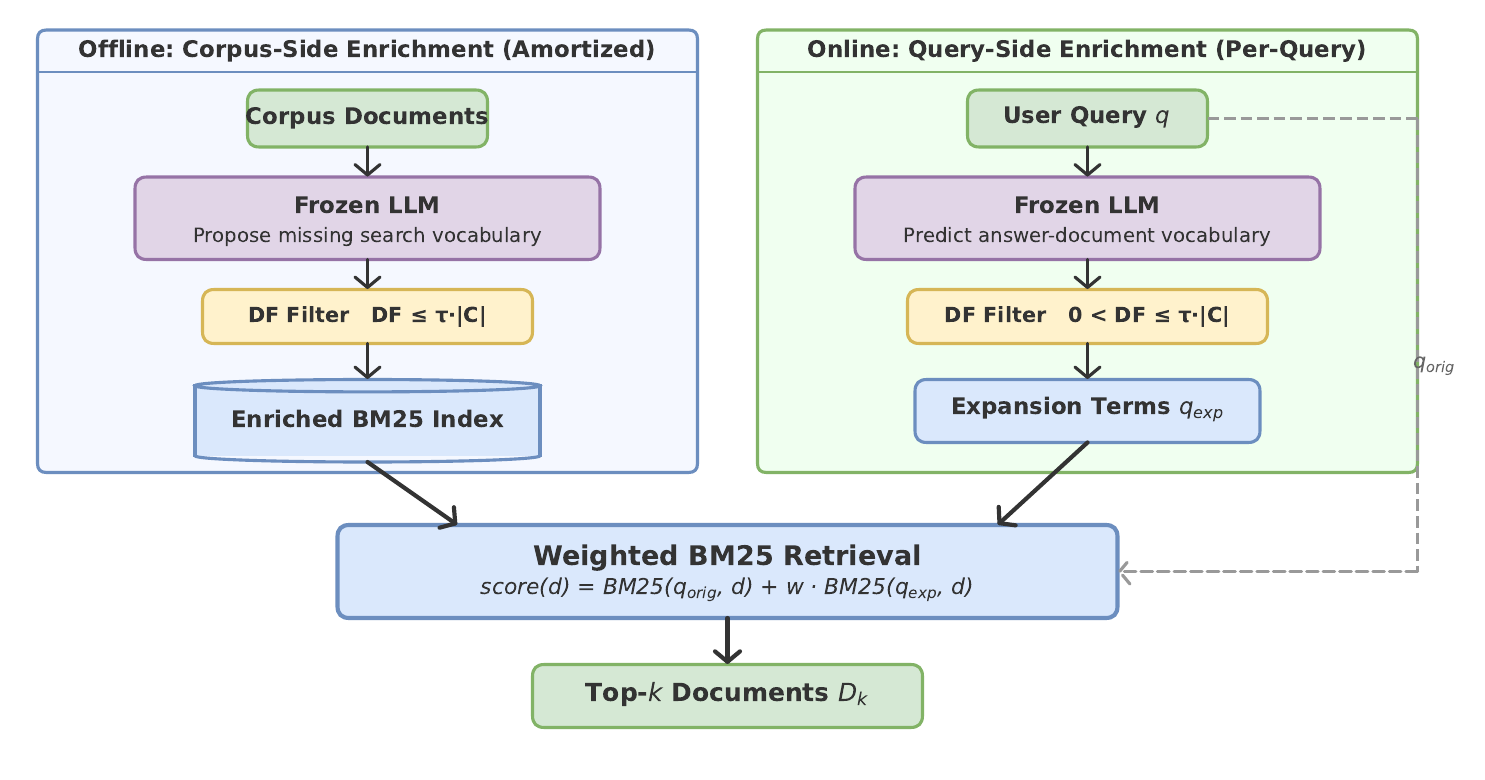}
\caption{\textbf{SIRA pipeline overview.}
Corpus-side enrichment (left) is performed once offline; query-side enrichment (right) runs per query.
Both stages apply a DF filter to reject uninformative terms; the query-side filter additionally requires $\text{DF} > 0$ to ensure each expansion term exists in the index.
The original query $q_{\text{orig}}$ bypasses enrichment (dashed) and is combined with the expansion terms in a single weighted BM25 call.}
\label{fig:sira_overview}
\end{figure}

\paragraph{How SIRA expertizes a new corpus.}
Given an unseen corpus, SIRA builds domain expertise from both the corpus side and the query side, as illustrated in \cref{fig:sira_overview}.
On the \emph{corpus side} (offline, once per corpus), the LLM reads each document, anticipates the search vocabulary a user would need to find it, and proposes candidate terms absent from the document text.
A document-frequency filter validates each candidate against the corpus index, discarding uninformative terms whose frequency exceeds an upper bound (detailed in \cref{sec:corpus_enrich}).
The surviving terms are injected into the BM25 index as atomic n-gram entries.
On the \emph{query side} (online, per query), the LLM first produces an \emph{expected-response sketch}: a compact set of concepts, entities, and discriminative terms likely to appear in a relevant document but absent from the query.
The same DF filter grounds the sketch in the enriched index.
Conditioned on the validated sketch, SIRA then compiles a \emph{retrieval program} and executes a single weighted BM25 call (detailed in \cref{sec:query_enrich}):
\begin{equation}
\text{score}(d) \;=\; \text{BM25}(q_{\text{orig}},\, d) \;+\; w \cdot \text{BM25}(q_{\text{exp}},\, d),
\label{eq:weighted_search}
\end{equation}
where $q_{\text{orig}}$ is the original query, $q_{\text{exp}}$ is the filtered expansion, and $w$ is the expansion weight. After the BM25 retrieval top-$k$ candidates are picked from the returned candidates based on relevance with the query.

\subsection{Vocabulary Enrichment with Superintelligent LLMs}
\label{sec:enrichment}
\label{sec:corpus_enrich}
\label{sec:query_enrich}

SIRA bridges the query--document vocabulary gap by enriching both sides with LLM-proposed terms. The goal of enrichment is not to add more text indiscriminately, but to surface compact lexical signals: concepts, aliases, and phrases that are likely to identify the desired evidence while remaining rare enough to distinguish it from the rest of the corpus.

This requires grounding LLM proposals in corpus statistics. A term that sounds expert-like may be useless for query-side enrichment if it never appears in the enriched index, and weak if it appears across many unrelated documents. SIRA therefore treats document frequency and BM25/TF--IDF-style salience as lightweight corpus-statistic tools: they check query-side term existence, estimate whether proposed terms are too common to discriminate, and retain terms that can contribute meaningful retrieval margin.

\paragraph{DF filter.}
\label{sec:df_filter}
To ensure that only corpus-grounded and discriminative terms enter the retrieval pipeline, both enrichment stages share a \emph{document-frequency (DF) filter}.
The filter enforces an upper bound $\text{DF} \leq \tau \cdot |C|$, pruning terms that are repeated across too much of the corpus and therefore receive little useful IDF weight.
For query-side enrichment, the filter additionally requires $\text{DF} > 0$, ensuring that every expansion phrase actually exists in the enriched index and can affect BM25 scoring.
Corpus-side enrichment does not require this lower bound, since enrichment itself introduces new vocabulary into the index.
The result is a compact set of terms that are plausible under the LLM's domain knowledge and measurable as useful search signals in the target corpus.

\paragraph{Corpus-side enrichment (offline).}
The goal is to anticipate how a user would search for a document when the vocabulary they would use is absent from the document text.
The prompt explicitly instructs the LLM to generate \emph{new} terms not already present in the document, focusing on discriminative vocabulary: synonyms, abbreviations, alternate names, and domain-specific phrasings that maximize lexical contrast with existing index terms.
Crucially, the prompt is task-aware: for claim-verification corpora, it emphasizes entity aliases and factual cues; for argument retrieval, it targets opposing-side vocabulary; for duplicate detection, it focuses on intent-preserving synonym substitutions.
Phrases that pass the DF filter are decomposed into sliding-window n-grams and injected into the corpus index as additional posting-list entries.

\paragraph{Query-side enrichment (online).}
The goal is the mirror image of corpus-side enrichment: predict vocabulary that a relevant answer document would use but that is absent from the query.
The prompt instructs the LLM to generate discriminative \emph{topic and domain vocabulary} that narrows the search space, while explicitly forbidding it from guessing the answer itself.
This distinction is critical for factoid queries, where predicting a named entity (e.g., a person or date) would bias retrieval toward a single candidate rather than broadening coverage of relevant evidence.
As on the corpus side, the prompt is task-aware: factoid QA targets contextual terms surrounding the answer, multi-hop QA distributes expansion across all entities and reasoning hops, and duplicate detection focuses on intent-preserving synonym substitutions.

\section{Experiments}
\label{sec:experiments}

We evaluate SIRA in three stages. First, we test pure retrieval quality on ten
BEIR benchmarks. Second, we ask whether this retrieval advantage transfers to
downstream question answering. Third, we evaluate hard search-agent retrieval on
BrowseComp-Wikipedia, a new BrowseComp-derived benchmark grounded in a
25,587,229-document English Wikipedia index.

\subsection{Experimental Setup}
\label{sec:setup}

We evaluate SIRA on ten BEIR datasets~\citep{thakur2021beir} spanning seven retrieval task types: question answering (NQ, HotpotQA), opinion retrieval (FIQA), fact-checking (FEVER, Climate-FEVER, SciFact), argument retrieval (ArguAna), citation prediction (SciDocs), and duplicate detection (Quora, CQADupStack).
\cref{tab:datasets} summarizes corpus sizes (5K to 5.4M documents) and the number of relevant documents per query (1.0 to 29.9), covering a broad range of retrieval regimes.
We report Recall@10 and NDCG@10.
Recall@10 measures the fraction of relevant documents appearing in the top-10 results, directly capturing a system's ability to surface evidence for a downstream reader.
NDCG@10 additionally rewards placing relevant documents at higher ranks, providing a complementary view of ranking quality.

BEIR is a particularly stringent testbed for retrieval agents because it evaluates retrieval directly rather than end-to-end answer generation. Many recent agentic search systems focus on QA-style benchmarks where a reader model, prior knowledge, or multi-round interaction can partially mask weak retrieval. In contrast, BEIR exposes the core retrieval problem: given a fixed corpus and a query, can the system surface the relevant documents in the top ranks? This makes it the right setting for evaluating SIRA as a retrieval agent rather than as a general QA agent.

\begin{table}[htb!]
  \caption{Overview of the ten BEIR retrieval benchmarks used in our evaluation, spanning diverse reasoning types including fact-checking, argumentation, citation prediction, and standard QA. The suite covers a broad range of domains and corpus sizes (5K to 5.4M documents), rigorously testing generalization. \textbf{Rel D/Q} denotes the average number of relevant documents per query.}
  \label{tab:datasets}
  \centering
  \small
  \resizebox{\textwidth}{!}{%
  \setlength{\tabcolsep}{4pt}
  \begin{tabular}{lccccl}
    \toprule
    \textbf{Dataset} & \textbf{Type} & \textbf{Queries} & \textbf{Corpus} & \textbf{Rel D/Q} & \textbf{Description} \\
    \midrule
    NQ & Question Answering & 3,452 & 2.68M & 1.22 & Retrieval for real-world Google search questions. \\
    HotpotQA & Question Answering & 7,405 & 5.23M & 2.00 & Multi-hop reasoning over Wikipedia paragraphs. \\
    FIQA & Opinion Retrieval & 648 & 57K & 2.63 & Financial QA over StackExchange data. \\
    ArguAna & Argument Retrieval & 1,401 & 8.67K & 1.00 & Matching counter-arguments for debate topics. \\
    CQADupStack & Duplicate Question & 1,570 & 40K & 2.40 & Duplicate detection across StackExchange forums. \\
    Quora & Duplicate Question & 10,000 & 523K & 1.57 & Duplicate detection for Quora questions. \\
    SciDocs & Citation Prediction & 1,000 & 26K & 29.93 & Predicting citations for scientific papers. \\
    FEVER & Fact-Checking & 6,666 & 5.42M & 1.19 & Verifying claims against Wikipedia text. \\
    Climate-FEVER & Fact-Checking & 1,535 & 5.42M & 3.05 & Verifying climate change claims. \\
    SciFact & Fact-Checking & 300 & 5K & 1.13 & Verifying scientific claims against abstracts. \\
    \bottomrule
  \end{tabular}}
\end{table}

\paragraph{Baselines.}
We compare against ten baselines spanning three retrieval paradigms.
BM25~\citep{robertson2009probabilistic} is the standard sparse lexical baseline.
Among neural methods, E5~\citep{wang2022text} is a dense bi-encoder trained on large-scale relevance data; SPLADE~\citep{formal2021splade} and SPARTA~\citep{zhao2021sparta} are learned sparse retrieval models that preserve inverted-index efficiency while learning term importance weights; Doc2Query~\citep{nogueira2019document} performs document expansion by predicting queries a document would answer.
Among LLM-based methods, HyDE~\citep{gao2023precise} generates a hypothetical document as a BM25 query expansion; CoT~\citep{wei2022chain} uses chain-of-thought prompting to expand the query with reasoning-derived terms; Search-R1~\citep{jin2025search} trains an RL policy for multi-round search; GrepRAG~\citep{wang2026greprag} generates grep-like pattern queries originally designed for code retrieval; and ShellAgent~\citep{subramanian2025keyword} relies on grep-based keyword-search tools within an agentic loop for multi-step retrieval.
HyDE, CoT, GrepRAG, ShellAgent, and SIRA use the same frozen LLM (Qwen3.6-35B-A3B-FP8); Search-R1 (E5) uses its publicly available trained checkpoint with the most potent E5 backend.
Neural baselines use official checkpoints (see~\cref{tab:main_results} caption).
SIRA uses Qwen3.6-35B-A3B-FP8 as its frozen LLM for both corpus-side and query-side enrichment.

\subsection{Main Results on Large-Scale Information Retrieval}
\label{sec:main_results}

In \cref{tab:main_results}, we report Recall@10 and NDCG@10 across all ten benchmarks, comparing SIRA against sparse, dense, and agentic baselines.
We structure the analysis around three research questions:
\begin{itemize}[leftmargin=*]
  \item \textbf{RQ1}: Can a training-free agentic system match or surpass trained sparse and dense retrievers?
  \item \textbf{RQ2}: Why do generic LLM search agents lag behind retrieval-native systems on BEIR, and does SIRA close this gap?
  \item \textbf{RQ3}: Does SIRA's retrieval quality translate to downstream QA performance competitive with RL-trained agentic QA systems?
\end{itemize}

\cref{tab:main_results} evaluates the central claim of this paper: a frozen LLM can turn BM25 from a lexical baseline into the strongest retriever in the comparison. SIRA uses the LLM not as a reader and not merely as a query-expansion module, but as a controller for the BM25 engine itself. It proposes missing vocabulary, grounds those proposals with corpus statistics, weights the surviving terms through BM25's IDF-sensitive scoring surface, and executes the result as a single ranked retrieval call.

This comparison is deliberately strict. The baselines include supervised dense and sparse retrievers trained on large-scale relevance data, as well as LLM-based query-expansion and search-agent baselines. Thus, the key question is not whether LLMs can generate plausible search text or perform more search rounds, but whether they can produce a corpus-discriminative retrieval action that ranks gold evidence above confusers.

\begin{table}[ht]
  \caption{Recall@10 and NDCG@10 on ten BEIR datasets. HyDE, CoT, GrepRAG, ShellAgent, and SIRA use Qwen3.6-35B-A3B-FP8 (frozen, 3B active parameters); Search-R1 (E5) uses its publicly available checkpoint with an E5 retrieval backend. Neural baselines use official checkpoints: SPARTA (BeIR/sparta-msmarco-distilbert-base-v1), SPLADE (naver/splade-cocondenser-ensembledistil), Doc2Query (doc2query/msmarco-t5-base-v1), E5 (intfloat/e5-base-v2). Best per dataset in \textbf{bold}; second best \underline{underlined}.}
  \label{tab:main_results}
  \centering
  \small
  \resizebox{\textwidth}{!}{%
  \setlength{\tabcolsep}{3.5pt}
  \begin{tabular}{l cccccccccc c}
    \toprule
    & ArguAna & C-FEVER & CQADup & FEVER & FIQA & HotpotQA & NQ & Quora & SciDocs & SciFact & Avg \\
    \midrule
    \multicolumn{12}{l}{\emph{Recall@10}} \\
    BM25 & .7738 & .1764 & .4163 & .6747 & .3198 & .6141 & .4543 & .9014 & .1636 & .8078 & .5302 \\
    \midrule
    Doc2Query & .7824 & .1761 & .4339 & .6769 & .3397 & .6527 & .5068 & .8975 & .1663 & .8270 & .5459 \\
    SPARTA & .6181 & .1050 & .3100 & .7246 & .2450 & .5356 & .5584 & .7445 & .1303 & .7084 & .4680 \\
    SPLADE & \underline{.8137} & .2881 & .4924 & .8954 & .4139 & .7027 & .7381 & .9206 & .1654 & .8230 & .6253 \\
    E5 & .7909 & .2899 & \underline{.5138} & \underline{.9109} & \underline{.4697} & \underline{.7276} & .7877 & \textbf{.9428} & .1962 & \underline{.8489} & \underline{.6478} \\
    \midrule
    HyDE & .7091 & .2598 & .3299 & .7132 & .2845 & .5051 & .4918 & .5151 & .1530 & .8344 & .4796 \\
    CoT & .7752 & .1867 & .4020 & .6765 & .3086 & .5789 & .4961 & .8704 & .1595 & .7961 & .5250 \\
    Search-R1 (E5) & .5760 & \underline{.3014} & .5008 & .9010 & .4499 & .6705 & \textbf{.7889} & .9355 & \underline{.2018} & .8349 & .6161 \\
    GrepRAG & .5746 & .0176 & .2400 & .1635 & .1009 & .3434 & .1628 & .5105 & .1027 & .5883 & .2804 \\
    ShellAgent & .2263 & .0305 & .2084 & .2557 & .1035 & .4327 & .1884 & .3330 & .0843 & .6685 & .2531 \\
    \midrule
    SIRA & \textbf{.9036} & \textbf{.3025} & \textbf{.6301} & \textbf{.9114} & \textbf{.4904} & \textbf{.7536} & \underline{.7883} & \underline{.9390} & \textbf{.2676} & \textbf{.9216} & \textbf{.6908} \\
    \midrule
    \midrule
    \multicolumn{12}{l}{\emph{NDCG@10}} \\
    BM25 & .4874 & .1372 & .3481 & .5036 & .2532 & .5851 & .2916 & .8055 & .1565 & .6791 & .4247 \\
    \midrule
    Doc2Query & .4946 & .1381 & .3667 & .5122 & .2731 & .6299 & .3302 & .7960 & .1589 & .6920 & .4392 \\
    SPARTA & .3890 & .0852 & .2497 & .6101 & .1925 & .5132 & .3983 & .6294 & .1272 & .5894 & .3784 \\
    SPLADE & .5253 & .2293 & .4083 & .7933 & .3478 & .6869 & .5369 & .8344 & .1586 & .7025 & .5223 \\
    E5 & \underline{.5323} & \underline{.2397} & \underline{.4196} & \underline{.8096} & \textbf{.3932} & \textbf{.6905} & \underline{.5835} & \textbf{.8648} & .1855 & \underline{.7156} & \underline{.5434} \\
    \midrule
    HyDE & .4366 & .2004 & .2463 & .5507 & .2223 & .4451 & .3315 & .3924 & .1402 & .6565 & .3622 \\
    CoT & .4951 & .1471 & .3354 & .4932 & .2486 & .5595 & .3168 & .7608 & .1518 & .6647 & .4173 \\
    Search-R1 (E5) & .3658 & \textbf{.2654} & .4040 & \textbf{.8215} & .3765 & .6520 & .5790 & \underline{.8543} & \underline{.1878} & .7094 & .5216 \\
    GrepRAG & .3555 & .0122 & .2048 & .0971 & .0763 & .2927 & .0908 & .4557 & .0921 & .4129 & .2090 \\
    ShellAgent & .1114 & .0206 & .1558 & .1298 & .0686 & .3417 & .1036 & .2476 & .0727 & .4374 & .1689 \\
    \midrule
    SIRA & \textbf{.6174} & .2288 & \textbf{.5327} & .8037 & \underline{.3771} & \underline{.6904} & \textbf{.5923} & .8490 & \textbf{.2449} & \textbf{.7866} & \textbf{.5723} \\
    \bottomrule
  \end{tabular}}
\end{table}

\paragraph{RQ1: SIRA surpasses trained retrievers without supervision.}
SIRA achieves the highest average Recall@10 on BEIR, reaching 0.691 compared with 0.648 for E5, 0.625 for SPLADE, and 0.530 for BM25. It does so without relevance labels, without fine-tuning a retriever, and without building an embedding index. The advantage also holds for ranking quality: SIRA reaches 0.572 average NDCG@10, compared with 0.543 for E5 and 0.522 for SPLADE. This is the central result: a training-free retrieval agent built on BM25 can outperform supervised dense and learned sparse retrievers trained on large-scale relevance data.

The gains are broad rather than driven by a single dataset. SIRA obtains the best Recall@10 on eight of ten benchmarks; the two exceptions are NQ, where Search-R1 (E5) is ahead by 0.06 percentage points, and Quora, where E5 is ahead by 0.4 percentage points. The largest improvements over E5 appear on datasets with structural query--document vocabulary gaps: +36\% relative on SciDocs, +23\% on CQADupStack, and +14\% on ArguAna. These are precisely the settings where corpus-grounded enrichment should help: the LLM proposes missing terminology, the DF filter removes absent or overly common terms, and BM25 amplifies the surviving discriminative vocabulary through IDF-weighted scoring.

\paragraph{RQ2: SIRA turns LLM reasoning into retrieval-native ranking.}
The LLM-based baselines remain below SIRA on pure retrieval metrics, but Search-R1 (E5) clarifies an important distinction. HyDE and CoT remain close to BM25 on average, with Recall@10 of 0.480 and 0.525 compared with 0.530 for BM25. Search-R1 gains substantially from its E5 backend, reaching 0.616 Recall@10 and 0.522 NDCG@10, but still trails SIRA's 0.691 Recall@10 and 0.572 NDCG@10. Stronger backend retrieval clearly helps, but a multi-round search policy on top of that backend still does not close the gap to corpus-grounded BM25 control.

The gap is even larger for grep-style tool-use agents. GrepRAG and ShellAgent use the same LLM backbone as SIRA, but treat retrieval as pattern generation rather than corpus-aware ranking, yielding average Recall@10 of only 0.280 and 0.253. Because the backbone is shared, the gap isolates the retrieval interface: grep-style agents search with patterns that lack BM25's document-frequency and IDF-weighted term scoring, while SIRA turns LLM proposals into weighted retrieval signals. SIRA therefore outperforms GrepRAG and ShellAgent by 41.0 and 43.8 absolute Recall@10 points.

Recent LLM search agents also show how much performance depends on the retrieval backend: Search-R1 improves markedly when paired with E5, yet still remains below E5 on average Recall@10 and below SIRA on both Recall@10 and NDCG@10. This does not mean E5 reasons better; it means retrieval-native systems are optimized for the object BEIR measures: ranking relevant documents above corpus-level distractors. SIRA closes this gap by using the LLM to program the retrieval engine itself: proposed terms are grounded by corpus statistics, filtered for discriminative value, weighted through BM25's IDF-sensitive scoring surface, and executed in a single ranked retrieval call.

\subsection{Downstream Question Answering}
\label{sec:downstream_qa}

Our goal is to test whether SIRA's retrieval advantage translates into downstream question answering. Since SIRA is a retrieval agent rather than an answer generator, the evaluation requires QA datasets with an associated retrieval corpus and gold evidence/answer annotations. Among the BEIR benchmarks, this leaves NQ and HotpotQA: both provide fixed corpora for retrieval and gold answer strings that allow us to measure whether retrieved documents contain the evidence needed to answer.

This setting directly tests the claim that a better retrieval agent can become a better QA agent by supplying stronger evidence to the reader. It is also a difficult comparison for SIRA. The baselines are recent agentic QA systems designed for these tasks, and many are trained or reinforced on QA-style objectives closely aligned with NQ and HotpotQA. They report end-to-end generated-answer accuracy, so they may receive credit even when retrieval is incomplete, because the reader can use parametric knowledge or answer synthesis. In contrast, SIRA is counted correct only when the gold answer string appears in the retrieved text.

We compare against six recent RL-trained agentic QA systems, including Search-R1~\citep{jin2025search}, TIPS~\citep{xie2026tips}, A$^2$Search~\citep{zhang20252}, E-GRPO~\citep{zhang2026grpo}, SSP~\citep{lu2025search}, and HiPRAG~\citep{wu2025hiprag}. We take the best reported numbers directly from the original papers, giving each baseline the benefit of its full end-to-end QA pipeline. SIRA uses no reader model in this comparison. It also uses no RL fine-tuning, no task-specific supervised training, and no multi-round search; the same one-shot SIRA retriever used in the BEIR experiments is evaluated directly. SIRA is evaluated only by \emph{answer coverage}, the fraction of queries for which its retrieved documents contain verifiable answer evidence.

\begin{figure}[t]
  \centering
    \vspace{-0.9cm}
\includegraphics[width=0.75\textwidth]{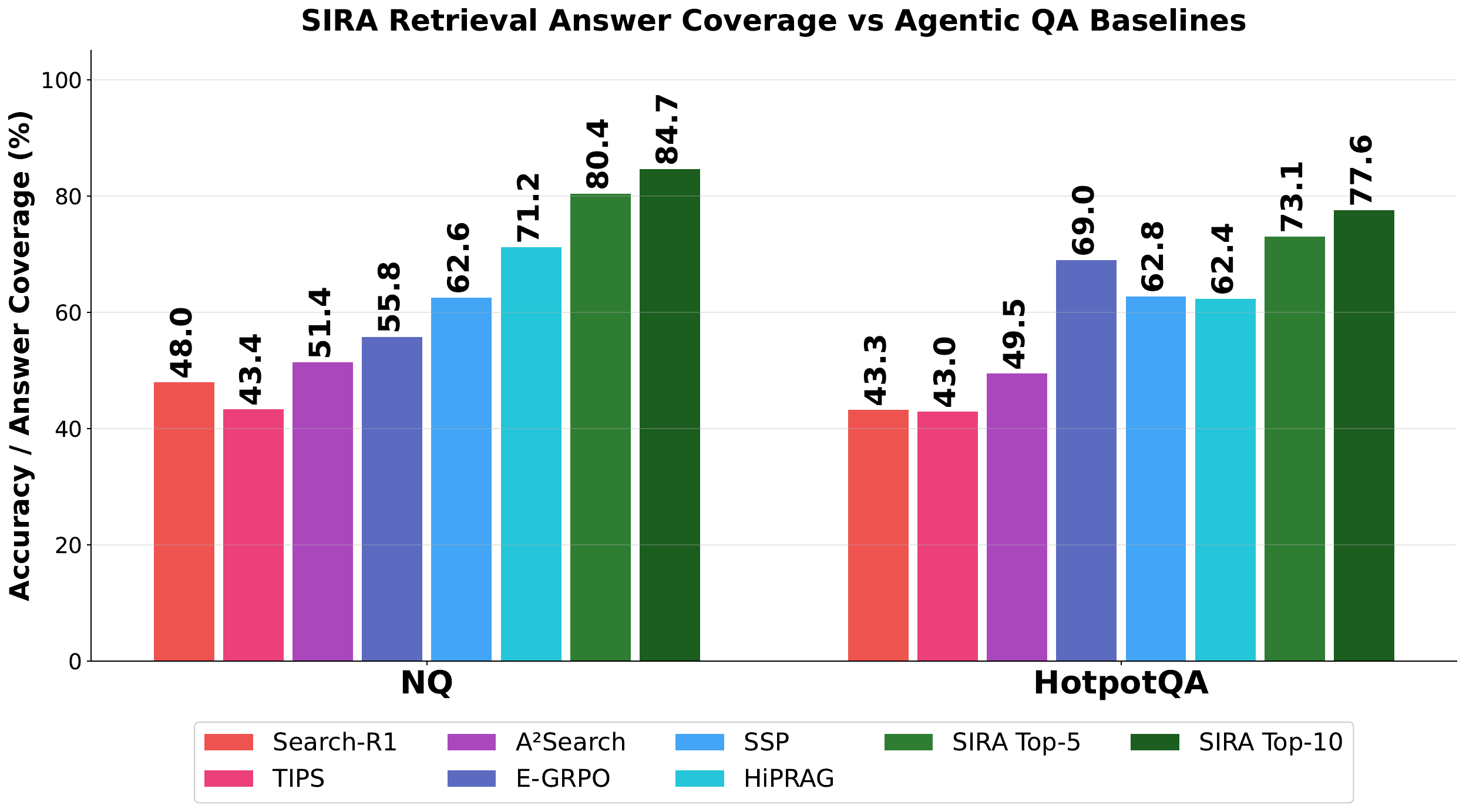}
  \caption{SIRA retrieval answer coverage (top-5 and top-10) vs.\ six RL-trained agentic QA systems on NQ and HotpotQA. Baseline numbers are the best reported results taken directly from each original paper. All baselines are end-to-end QA pipelines reporting generated-answer accuracy; SIRA is a pure retriever with no reader. Answer coverage requires the gold string to be retrieved.}
  \label{fig:sira_vs_searchr1}
  \vspace{-0.5cm}
\end{figure}

\paragraph{RQ3: The best retrieval agent yields the strongest QA evidence.}
Despite this disadvantaged evaluation, SIRA's retrieval-only answer coverage exceeds the reported end-to-end QA accuracy of all six agentic baselines at top-10. This comparison gives the baselines both their trained search policies and their answer generators, while SIRA contributes only retrieved text. As shown in \cref{fig:sira_vs_searchr1}, SIRA reaches 84.7\% on NQ and 77.6\% on HotpotQA. The strongest baselines: HiPRAG reaches 71.2\% on NQ, while E-GRPO reaches 69.0\% on HotpotQA. SIRA exceeds both by large margins.

The result is also strong at a tighter retrieval budget. At top-5, SIRA achieves 80.4\% on NQ and 73.1\% on HotpotQA, still exceeding every baseline in the comparison. These results support a simple conclusion: for corpus-grounded QA, improving retrieval can be more important than adding more search rounds or training the answer generator. The compared agents are built for end-to-end QA and many are optimized on QA-style rewards, yet SIRA surfaces answer-bearing passages more reliably with a single retrieval call.

\subsection{BrowseComp-Wikipedia: Hard Search at Wikipedia Scale}
\label{sec:browsecomp_wikipedia}

BrowseComp was introduced as a benchmark for evaluating browsing agents on
hard-to-find, short-answer questions that require nontrivial search over the
web~\citep{wei2025browsecomp}. The benchmark is valuable because the answers are easy
to verify once found, while the evidence-finding process can require substantial
exploration. However, for evaluating search itself, the original BrowseComp
setting has two important limitations. First, it depends on live web search, so
the searchable corpus, ranking function, snippets, and availability of evidence
are controlled by an external and changing system. Second, as BrowseComp has
become a widely used benchmark for deep-research and browsing agents, the
questions and answers are increasingly exposed through papers, leaderboards,
logs, and repeated evaluations, making leakage and overfitting a growing
concern.

A recent line of work has tried to make BrowseComp-style evaluation more
reproducible by converting it into a fixed-corpus setting. BrowseComp-Plus, for
example, contains 830 queries over a fixed corpus of 100,195
documents~\citep{chen2025browsecomp}. This is an important step toward controlled
retrieval evaluation, but the corpus is still small relative to the setting in
which search agents are normally expected to operate. In a corpus of roughly one
hundred thousand documents, the search problem can be made easier by
benchmark-specific engineering, memorization, or overfitting to a relatively
limited evidence pool. This is precisely the issue we argue against throughout
the paper: the central difficulty of search is not merely producing a
semantically plausible query, but disambiguating the right page from a very large
number of plausible confusers. That difficulty becomes much more realistic when
the search space contains millions of pages rather than only a small curated
corpus.

To create a harder and more realistic evaluation, we construct
\textbf{BrowseComp-Wikipedia}, a new benchmark that combines BrowseComp-style
query difficulty with Wikipedia-scale retrieval. We identify 232 BrowseComp
queries whose answers can be fully supported using English Wikipedia pages
alone. For each query, we annotate the gold Wikipedia URL or URLs that contain
sufficient evidence to answer the question. The retrieval corpus is an indexed
English Wikipedia collection containing \textbf{25,587,229 documents}. This
creates a substantially larger and more natural retrieval setting than existing
fixed-corpus BrowseComp variants: the queries retain the hard, indirect,
entity-seeking structure of BrowseComp, but the search system must recover the
answer-bearing page from a corpus with tens of millions of Wikipedia documents.

This benchmark fills an important gap in search-agent evaluation. Original
BrowseComp evaluates end-to-end browsing over the live web. BrowseComp-Plus
provides a controlled but relatively small fixed-corpus benchmark.
BrowseComp-Wikipedia instead evaluates whether a search agent can retrieve the
correct Wikipedia page from a large, public, reproducible corpus. The task is
deliberately strict: success requires retrieving the gold page that can support
the answer, not merely generating the correct final answer from parametric memory
or from indirectly related evidence.

\paragraph{SIRA without index-time LLM enrichment.}
For this experiment, we evaluate a deliberately toned-down version of SIRA. In
the main method, SIRA uses index-time enrichment to add missing search vocabulary
to documents. Here, we remove that expensive LLM-based document enrichment step.
Instead, we use Wikipedia's own category system as a natural source of
corpus-side enrichment. Each Wikipedia document is indexed using its text,
unigrams, bigrams, and category strings. Category strings are treated as
protected lexical features: for example, a category such as
\texttt{1927 births} is indexed as an atomic feature such as
\texttt{category:1927\_births}, rather than being split into unrelated terms.

This category information is useful because many BrowseComp-style questions
describe the target entity indirectly. A question may not name the answer
directly, but it may imply constraints such as a birth year, death year,
occupation, nationality, award, genre, institution, historical period, or place.
Wikipedia categories expose many of these attributes in a corpus-visible form.
Examples include categories such as \texttt{1927 births},
\texttt{American film actresses}, \texttt{British science fiction writers},
\texttt{Nobel laureates in Physics}, and
\texttt{People educated at Eton College}. These categories give SIRA precise
retrieval handles for attributes that may be only indirectly expressed in the
original query.

To prevent the LLM from hallucinating category constraints, SIRA is not allowed
to freely invent category tokens. During query construction, the LLM may propose
potentially useful Wikipedia categories, but every proposed category must be
grounded through the Wikipedia category graph before it can be used in the final
BM25 query. The category-graph tool verifies whether the category exists and can
return valid neighboring categories when appropriate. Invalid or unsupported
categories are discarded. Thus, the LLM contributes semantic search planning, but
the final retrieval query remains grounded in corpus-observable structure.

For BrowseComp-Wikipedia, SIRA is not single-query unlike in the easier strict BEIR sense. Instead, it is evidence-blind and budgeted: the agent may issue multiple
category-grounded BM25 calls, but every returned page consumes the fixed
retrieval budget, and recall is measured over the resulting stream of unique
Wikipedia pages. SIRA may fetch titles/snippets only for pages already inside the fixed page budget, and fetched content is used only for final ranking, not for formulating subsequent BM25 queries.

\paragraph{Agentic search with the Perplexity search engine.}
As strong agentic baselines, we equip each frozen backbone LLM with the same
production web-search engine and allow it to browse iteratively. Concretely, the agent is given a single retrieval tool, \texttt{perplexity\_search\_wikipedia\_en}, which performs a Perplexity-based web search and returns a ranked list of up to 20
results per query. Each result is returned as structured output containing the
title, URL, and snippet, which the agent can read before deciding what to do
next. To match the evidence space available to SIRA, the search tool is
constrained to \texttt{en.wikipedia.org}: all other pages, including non-English
Wikipedia subdomains, are filtered out. This gives the agent a clean,
single-source evidence stream from English Wikipedia.

The Perplexity baseline follows the standard BrowseComp evaluation protocol with
the default prompt. The agent can issue searches, inspect returned titles and
snippets, reformulate its query, and continue browsing until it commits to an
answer, for up to 100 turns per question. This baseline represents the prevailing
multi-round search-agent paradigm: an LLM controls a commercial search engine
through repeated query reformulation and intermediate result inspection. In
contrast, SIRA represents a corpus-grounded retrieval paradigm: the LLM must
compile a retrieval query against the indexed corpus before seeing retrieved
pages.

\paragraph{Evaluation protocol.}
Our evaluation protocol is intentionally strict for SIRA. The LLM is allowed to
think as much as needed before issuing a BM25 query, and it may use the
category-graph tool to validate candidate category constraints. The prompt used
for SIRA is shown in Figure~\ref{fig:sira_browsecomp_prompt}. However, once the
BM25 query is fired, every returned page consumes retrieval budget. SIRA is not
allowed to inspect retrieved snippets or pages and then reformulate the query
using that retrieved context. This differs from the Perplexity baseline, which
can read intermediate search results and use them to issue better follow-up
queries.

We evaluate retrieval budgets of $B \in \{1, 10, 100\}$ pages. A query is
counted as successful if at least one gold Wikipedia URL appears among the pages
returned within the budget. For SIRA, this corresponds directly to the top
$B$ BM25 results. For the Perplexity baseline, we consider the ordered stream of
unique English Wikipedia URLs surfaced to the agent during its iterative browsing
process and truncate that stream at the same budgets. This allows us to measure
gold-page recall under comparable evidence budgets, even though the baseline is
allowed to browse interactively.

The resulting metric is gold-page recall: the fraction of the 232 queries for
which at least one annotated gold Wikipedia page is retrieved within the page
budget. Formally,
\[
\mathrm{Recall@}B
=
\frac{1}{|\mathcal{Q}|}
\sum_{q \in \mathcal{Q}}
\mathbf{1}
\left[
G_q \cap R_B(q) \neq \emptyset
\right],
\]
where $\mathcal{Q}$ is the set of BrowseComp-Wikipedia queries, $G_q$ is the set
of gold Wikipedia URLs for query $q$, and $R_B(q)$ is the set of at most $B$
retrieved Wikipedia URLs. Recall@1 measures whether the system ranks the
answer-bearing page first. Recall@10 measures whether the answer page fits
within a small reader context. Recall@100 measures whether the retriever can
recover the target page under a larger but still finite retrieval budget.

\paragraph{Results.}
The results are summarized in Table~\ref{tab:browsecomp_wiki_results}. The
strongest SIRA configuration achieves the best recall at every retrieval budget,
despite operating under a substantially stricter protocol than the agentic
search baselines built on the Perplexity search engine. With Claude 4.6 Opus as the backbone, SIRA obtains
9.70\% Recall@1, 15.27\% Recall@10, and 36.14\% Recall@100. Compared with the
best Perplexity baseline at each budget, this corresponds to gains of
6.68 percentage points at Recall@1, 8.37 percentage points at Recall@10, and
3.81 percentage points at Recall@100.

The gains are largest in the low-budget regime, which is the most important
setting for retrieval-augmented agents. Recall@1 measures whether the system
places the answer-bearing page first, and Recall@10 measures whether the page is
available within a small reader context. SIRA more than triples the best
Perplexity Recall@1 score and more than doubles the best Perplexity Recall@10
score. This is notable because the Perplexity-search agents are allowed to browse
interactively for up to 100 turns, reading titles and snippets before issuing
follow-up searches. SIRA, by contrast, must commit to a single corpus-grounded
BM25 query before reading any retrieved page.

\begin{table}[t]
\centering
\small
\setlength{\tabcolsep}{5.5pt}
\renewcommand{\arraystretch}{1.08}
\begin{tabular}{@{}llccc@{}}
\toprule
\textbf{System}
& \textbf{Backbone LLM}
& \textbf{Recall@1}
& \textbf{Recall@10}
& \textbf{Recall@100} \\
\midrule

\multicolumn{5}{@{}l}{\emph{Agentic baseline: Perplexity search over English Wikipedia, up to 100 turns}} \\
\quad Perplexity
& Claude 4.6 Opus
& \underline{2.59}
& \underline{4.74}
& \underline{32.33} \\
\quad Perplexity
& GPT-5.4
& \underline{3.02}
& \underline{6.90}
& \textbf{31.47} \\

\addlinespace[2pt]
\midrule

\multicolumn{5}{@{}l}{\emph{SIRA: single-shot, category-grounded BM25 on
 Wikipedia}} \\
\quad SIRA
& Claude 4.6 Opus
& \textbf{9.70}
& \textbf{15.27}
& \textbf{36.14} \\
\quad SIRA
& GPT-5.4
& \textbf{5.71}
& \textbf{13.13}
& \underline{18.51} \\

\bottomrule
\end{tabular}
\caption{
Gold-page recall on BrowseComp-Wikipedia. All values are percentages. The
benchmark contains 232 BrowseComp queries whose answers can be fully supported
using English Wikipedia. The retrieval corpus contains 25,587,229 indexed
Wikipedia documents. SIRA uses no index-time LLM enrichment in this experiment;
instead, it uses Wikipedia categories as natural corpus-side enrichment and
validates proposed category tokens against the Wikipedia category graph. The
Perplexity baseline browses interactively over \texttt{en.wikipedia.org} for up
to 100 turns, while SIRA must issue corpus-grounded BM25 queries before
reading any retrieved page. For each backbone LLM, the better of the two search
engines (Perplexity vs.\ SIRA) is shown in \textbf{bold} and the other is
\underline{underlined}.
}
\label{tab:browsecomp_wiki_results}
\end{table}

These results support the central hypothesis of this paper: hard search is not
only about having a powerful LLM or a strong commercial search engine, but about
constructing the right corpus-grounded retrieval action. The Perplexity-search agents
represent the prevailing multi-round paradigm, where an LLM repeatedly searches,
reads partial results, and reformulates. SIRA instead uses the LLM to compile a
precise retrieval program before evidence consumption. Its strong low-budget
recall shows that Wikipedia categories provide discriminative, corpus-visible
handles that allow SIRA to rank the correct page above millions of plausible
confusers.

Overall, BrowseComp-Wikipedia provides a new controlled benchmark for hard
search-agent evaluation. It preserves the difficulty of BrowseComp-style
queries, but grounds evaluation in a public Wikipedia corpus with 25,587,229
indexed documents and gold URL supervision. It also separates two search
paradigms: iterative web-search agents that rely on reading intermediate
results, and corpus-grounded retrieval agents that must construct a precise
search action before consuming evidence. The strong performance of SIRA in this
setting supports the central claim of the paper: hard search is not only about
semantic understanding, but about constructing grounded retrieval programs that
can disambiguate the right page from a very large corpus.

\paragraph{Limitations and future work.}
SIRA assumes that the frozen LLM can understand the query and provide useful semantic priors about the target corpus. We have not evaluated settings where the corpus is far outside the LLM's pretraining distribution; in such domains, corpus-side adaptation or fine-tuning may be needed before the LLM can propose reliable enrichment terms.

\section{Conclusion}
\label{sec:conclusion}

We introduced SIRA, a retrieval-centric agent that turns LLM reasoning into
controllable lexical retrieval. Instead of using the LLM to repeatedly query a
black-box search tool, read snippets, and reformulate, SIRA uses the LLM to
program the retrieval action itself: enrich the corpus with missing user
vocabulary, enrich the query with likely evidence vocabulary, validate proposed
terms with corpus statistics, and execute a single weighted BM25 call.

The main result is that this simple interface changes the role of BM25. Across
ten BEIR benchmarks, SIRA achieves the highest average Recall@10 and NDCG@10 in
our comparison, outperforming BM25, E5, SPLADE, and recent LLM-based search
agents while using no relevance labels, no retriever fine-tuning, and no
embedding index. Its gains are broad, with the best Recall@10 on eight of ten
datasets and especially large improvements on tasks where query and document
vocabularies diverge.

The downstream QA results show that this retrieval advantage matters beyond
standard IR metrics. On NQ and HotpotQA, SIRA's retrieval-only answer coverage
exceeds recent RL-trained agentic QA systems, even though those systems use
trained search policies and answer generators while SIRA contributes only
retrieved evidence. This supports a simple conclusion: for corpus-grounded QA,
the ability to surface the right evidence can matter more than adding more
search rounds.

Finally, BrowseComp-Wikipedia shows that the same principle extends to hard
agentic search at much larger scale. We construct a 232-query benchmark from
BrowseComp whose answers can be supported entirely by English Wikipedia, and
evaluate retrieval over a 25,587,229-document Wikipedia index. In this setting,
we remove SIRA's index-time LLM document enrichment and use only Wikipedia
categories as natural corpus-side structure, grounding every proposed category
through the Wikipedia category graph. Even under this stripped-down protocol,
SIRA outperforms Perplexity-based agents at every retrieval budget, despite the
baselines being allowed to browse iteratively for up to 100 turns. The strongest
SIRA configuration reaches 9.70\% Recall@1, 15.27\% Recall@10, and 36.14\%
Recall@100.

Together, these results suggest a different path for retrieval-augmented agents.
Rather than making agents search longer, accumulate more context, and rely on
increasingly expensive interaction with black-box search engines, we can make
the retrieval action itself more expert, corpus-aware, and interpretable. The
remaining open question is how far this idea extends to corpora far outside the
frozen LLM's knowledge, where corpus adaptation or fine-tuning may be needed
before reliable enrichment is possible. But at least for broad public knowledge
corpora and standard retrieval benchmarks, SIRA shows that one well-grounded
retrieval program can beat many rounds of exploratory search.

\clearpage
\newpage
\bibliographystyle{assets/plainnat}
\bibliography{refs}

\clearpage
\newpage
\beginappendix

\section{BrowseComp-Wikipedia Search Prompt}
\label{app:browsecomp_prompt}

For the BrowseComp-Wikipedia experiment (\cref{sec:browsecomp_wikipedia}), SIRA is driven by the single prompt shown in \cref{fig:sira_browsecomp_prompt}.
The prompt instructs the LLM to first solve the clue chain privately, validate candidate Wikipedia categories through the \texttt{search\_categories} tool, and then commit to one or more grounded \texttt{bm25\_search} calls under a fixed 100-unique-page retrieval budget.
The template variable \texttt{\{problem\}} is replaced by the BrowseComp question at evaluation time.

\begin{tcolorbox}[colback=gray!5, colframe=gray!50, fontupper=\small\ttfamily, breakable]
You are a budgeted Wikipedia BM25 search agent for hard clue-chain questions. Your goal is to put the gold page in the final top 10 or top 100 using at most 100 unique BM25-retrieved pages.\\[4pt]
TOOLS:\\[4pt]
\hspace*{1em}search\_categories(query: str, k: int = 20) -> str\\
\hspace*{2em}Search real Wikipedia category names. Use this BEFORE bm25\_search to find closest real category anchors for your intended entity/type/geography facet. If the exact imagined category is absent, broaden the query and choose the closest semantically aligned real category returned by the tool.\\[4pt]
\hspace*{1em}bm25\_search(keywords: str, categories: list[str] = [], k: int = 10) -> str\\
\hspace*{2em}Runs BM25 and spends the fixed 100-unique-page retrieval budget. Returns only new pages. Call only when keywords are concrete and categories are real/validated. Keep issuing targeted BM25 calls until this returns BUDGET\_EXHAUSTED.\\[4pt]
\hspace*{1em}fetch\_page(page\_id: int) -> str\\
\hspace*{2em}Fetch title + snippet for a page already returned by bm25\_search. Use this only to rank or verify candidates already inside the 100-page budget.\\[4pt]
WORKFLOW:\\
\hspace*{1em}1. Privately solve the puzzle as far as possible. Identify the likely answer type, answer hypotheses, clue entities, bridge entities, locations, dates, works, organizations, and relationships.\\
\hspace*{1em}2. Before every bm25\_search, validate categories. Call search\_categories with broad and narrow variants until you find the closest real category anchors for that query facet. Do not invent categories. Do not leave categories empty merely because the exact category is missing; pick the closest real semantically aligned category returned by the tool.\\
\hspace*{1em}3. Gate each BM25 query. Fire bm25\_search only if the keywords contain concrete Wikipedia-surface strings and the categories are real and aligned with the same hypothesis/facet. If categories are broad but aligned, use them with strong rare keywords. If keywords are generic or the category points to a different domain, do not spend budget on that query.\\
\hspace*{1em}4. Use multiple targeted BM25 queries until the 100-page budget is exhausted. Do not stop merely because you found a plausible candidate; continue spending the budget on distinct high-confidence facets so the final 100 pages have the best possible recall. Good query facets include: final answer hypothesis, bridge entity, clue entity, event/work/institution facet, and category/type/geography facet.\\
\hspace*{1em}5. Track all page\_ids returned by bm25\_search. Optionally fetch the most promising candidates to rank them, but keep querying until bm25\_search returns BUDGET\_EXHAUSTED or unique\_pages\_seen reaches 100. Final page\_ids must come from pages actually seen in bm25\_search/fetch\_page results.\\[4pt]
FINAL OUTPUT:\\
Output ONLY JSON with exactly these fields:\\[4pt]
\hspace*{1em}"page\_ids" --- ordered list of up to 100 page\_ids, most likely gold first. Use only page\_ids seen in this session.\\[4pt]
\hspace*{1em}"rationale" --- one short paragraph explaining the ranking evidence.\\[4pt]
\hspace*{1em}"hypothesis" --- one compact sentence naming the likely answer and key clue entities.\\[4pt]
Do not output markdown fences or prose outside JSON.\\[4pt]
Question:\\
\{problem\}
\end{tcolorbox}
\captionof{figure}{Prompt used to drive SIRA on the BrowseComp-Wikipedia benchmark. The LLM reasons privately, grounds candidate Wikipedia categories through the \texttt{search\_categories} tool, and issues grounded \texttt{bm25\_search} calls under a fixed 100-unique-page budget. The variable \texttt{\{problem\}} is replaced by the BrowseComp question.}
\label{fig:sira_browsecomp_prompt}

\section{BrowseComp-Wikipedia Case Study: SIRA vs.\ Perplexity}
\label{app:browsecomp_case_study}

\cref{fig:browsecomp_case_study} shows the complete tool-call traces for a single BrowseComp-Wikipedia query (gold page: \emph{LocoRoco}).
SIRA reaches the gold page at rank~1 after a handful of category-grounded BM25 calls, while the agentic Perplexity-search baseline issues 72 distinct searches, never retrieves the gold page, and commits to an incorrect answer.

\begin{tcolorbox}[colback=gray!5, colframe=gray!50, fontupper=\footnotesize\ttfamily, breakable]
\textbf{Question} (paraphrased to avoid benchmark leakage; gold page: LocoRoco, page id 4905353):\\
A certain video game made its debut on a gaming platform in July of a year somewhere from 2004 through 2007. The person who created it was born in April of a year in the 1970--1973 range and drew the game's earliest sketches at some point between 2002 and 2005. One of the title's music composers was born in June of a year from 1968 to 1971. Which game is being described?\\[6pt]
\textbf{SIRA (ours) --- gold retrieved at rank 1.} 11 bm25\_search calls, 9 search\_categories calls, 9 fetch\_page calls; budget not exhausted. Trace ($\star$ = call whose results contained the gold page):\\[3pt]
\hspace*{1em}0. [search\_categories] query: video games released July 2005 2006\\
\hspace*{1em}1. [search\_categories] query: Japanese video game designers born 1970s\\
\hspace*{1em}2. [search\_categories] query: video games 2005 2006 2007\\
\hspace*{1em}3. [search\_categories] query: indie video games\\
$\star$\hspace*{0.5em}4. [bm25\_search] keywords: LocoRoco game PSP 2006 | categories: ['2006 video games'] | k=5\\
\hspace*{1em}5. [search\_categories] query: PlayStation Portable games\\
\hspace*{1em}6. [fetch\_page] page\_id: 4905353\\
\hspace*{1em}7. [bm25\_search] keywords: Tsutomu Kouno LocoRoco designer creator | categories: ['Japanese video game designers'] | k=5\\
\hspace*{1em}8. [search\_categories] query: video game composers born 1968 1969 1970\\
\hspace*{1em}9. [search\_categories] query: video game composers\\
\hspace*{1em}10. [search\_categories] query: video game music composers Japanese\\
\hspace*{1em}11. [bm25\_search] keywords: Fumito Ueda game designer born April 1970 | categories: ['Japanese video game designers'] | k=5\\
\hspace*{1em}12. [fetch\_page] page\_id: 3041309\\
\hspace*{1em}13. [bm25\_search] keywords: video game released July 2005 sketches development | categories: ['2005 video games'] | k=10\\
\hspace*{1em}14. [bm25\_search] keywords: Kow Otani composer born June | categories: ['Japanese video game composers'] | k=5\\
\hspace*{1em}15. [search\_categories] query: Shadow of the Colossus\\
\hspace*{1em}16. [fetch\_page] page\_id: 3402872\\
\hspace*{1em}17. [bm25\_search] keywords: Osu Tatakae Ouendan game July 2005 | categories: ['2005 video games'] | k=5\\
\hspace*{1em}18. [fetch\_page] page\_id: 2740521\\
\hspace*{1em}19. [bm25\_search] keywords: Keiichi Yano video game designer iNiS born | categories: ['Japanese video game designers'] | k=5\\
\hspace*{1em}20. [fetch\_page] page\_id: 20435930\\
\hspace*{1em}21. [bm25\_search] keywords: game designer born April 1971 creator sketches | categories: ['Japanese video game designers'] | k=5\\
\hspace*{1em}22. [bm25\_search] keywords: game designer born April 1972 creator | categories: ['Japanese video game designers'] | k=5\\
\hspace*{1em}23. [fetch\_page] page\_id: 328218\\
\hspace*{1em}24. [fetch\_page] page\_id: 5262474\\
\hspace*{1em}25. [bm25\_search] keywords: Shinta Nojiri Metal Gear Acid game director July | categories: ['2005 video games', '2006 video games'] | k=5\\
\hspace*{1em}26. [fetch\_page] page\_id: 851696\\
\hspace*{1em}27. [fetch\_page] page\_id: 1918270\\
\hspace*{1em}28. [bm25\_search] keywords: Tsutomu Kouno LocoRoco creator born | categories: ['Japanese video game designers'] | k=5\\[6pt]
\textbf{Perplexity --- 72 searches, gold page never retrieved; final answer incorrect.} Query trace:\\[3pt]
\hspace*{1em}1. video game creator born April 1970 1971 1972 1973 first sketches 2002 2003 2004 2005\\
\hspace*{1em}2. video game debuted July 2004 2005 2006 2007 platform\\
\hspace*{1em}3. Painkiller video game 2004 release date platform July\\
\hspace*{1em}4. Adrian Chmielarz Painkiller game development sketches history\\
\hspace*{1em}5. Painkiller video game Xbox release date July 2006 music composer\\
\hspace*{1em}6. Painkiller video game soundtrack composer born June 1968 1969 1970 1971\\
\hspace*{1em}7. Shoji Meguro video game composer games list Persona\\
\hspace*{1em}8. Painkiller video game full Wikipedia article music composer development\\
\hspace*{1em}9. Persona 3 video game release date July 2006 director creator\\
\hspace*{1em}10. Trauma Center Under the Knife release date July 2004 2005 director creator born April\\
\hspace*{1em}11. Katsura Hashino born birthday Persona 3 director development history\\
\hspace*{1em}12. Persona 3 development history first sketches concept 2002 2003 2004\\
\hspace*{1em}13. Katsura Hashino birthday born date April 1970 1971 1972 1973\\
\hspace*{1em}14. Persona 3 Wikipedia full article director producer composer\\
\hspace*{1em}15. Katsura Hashino Atlus director born 1970s April\\
\hspace*{1em}16. Persona 3 game creator director Hashino birth date Wikipedia\\
\hspace*{1em}17. video game released July 2004 2005 2006 2007 creator born April 1970-1973 composer born June 1968-1971\\
\hspace*{1em}18. Maken X Atlus 1999 director Katsura Hashino born April\\
\hspace*{1em}19. Persona 3 Wikipedia infobox director producer release date\\
\hspace*{1em}20. Todd Howard born 1970 April game director Bethesda first sketches 2002 2003 2004 2005\\
\hspace*{1em}21. The Elder Scrolls IV Oblivion release date July 2006 development first sketches\\
\hspace*{1em}22. Todd Howard born April 1970 birthday Wikipedia\\
\hspace*{1em}23. video game composer born June 1968 1969 1970 1971 games released July 2004 2005 2006 2007\\
\hspace*{1em}24. Persona 3 development "first sketches" OR "concept art" OR "early development" 2003 2004\\
\hspace*{1em}25. killer7 video game July 2005 director creator Suda51 born April\\
\hspace*{1em}26. Suda51 Goichi Suda born April 1968-1973 game director\\
\hspace*{1em}27. killer7 music composer born June 1968 1969 1970 1971\\
\hspace*{1em}28. Hidetaka Suehiro SWERY games directed 2004 2005 2006 2007 release July\\
\hspace*{1em}29. The Idolmaster video game 2005 arcade director creator born April composer born June\\
\hspace*{1em}30. Persona 3 Wikipedia "first sketches" OR "initial sketches" development 2003\\
\hspace*{1em}31. Akihiro Ishihara Idolmaster director born April 1970 1971 1972 1973\\
\hspace*{1em}32. video game Wikipedia "first sketches" development creator director born April\\
\hspace*{1em}33. The Idolmaster video game Wikipedia development history creator Ishihara first sketches\\
\hspace*{1em}34. Jason Jones Bungie game director Halo released July 2004 2005 2006 2007\\
\hspace*{1em}35. game released July 2004-2007 creator born April 1970-1973 "first sketches"\\
\hspace*{1em}36. Katsura Hashino Wikipedia personal life born 1970 1971 1972 1973\\
\hspace*{1em}37. Michel Ancel born April 1972 Beyond Good Evil game development sketches\\
\hspace*{1em}38. video game Wikipedia "first sketches" development 2002 2003 2004 2005 released July\\
\hspace*{1em}39. video game Wikipedia "first sketches" 2002 2003 2004 2005 creator born April\\
\hspace*{1em}40. Shantae game creator Matt Bozon born April 1970 1971 1972 1973 first sketches\\
\hspace*{1em}41. Viva Piñata video game release date July 2006 Gregg Mayles development first sketches\\
\hspace*{1em}42. Gregg Mayles Rare games 2004 2005 2006 2007 director creator Viva Pinata\\
\hspace*{1em}43. Grant Kirkhope born June 1968 1969 1970 1971 composer birthday\\
\hspace*{1em}44. Viva Piñata video game release date November 2006 Xbox 360\\
\hspace*{1em}45. video game composer born June 1968 games released July 2004 2005 2006 2007\\
\hspace*{1em}46. Persona 3 Wikipedia full article infobox director producer release date July 13 2006\\
\hspace*{1em}47. Motoi Sakuraba born June 1968 1969 1970 1971 composer Tales of Symphonia\\
\hspace*{1em}48. Tales of Symphonia director creator born April 1970-1973 development first sketches\\
\hspace*{1em}49. Prey 2006 video game director creator born April 1970-1973 development first sketches composer\\
\hspace*{1em}50. Persona 3 Wikipedia Katsura Hashino born April 1970 1971 1972 1973\\
\hspace*{1em}51. \={O}kami video game release date July 2006 director Hideki Kamiya born April composer\\
\hspace*{1em}52. Hideki Kamiya born April 1970 1971 1972 1973 birthday\\
\hspace*{1em}53. Shin Megami Tensei Digital Devil Saga release date July 2004 director Katsura Hashino\\
\hspace*{1em}54. video game released July 2004-2007 composer born June 1968-1971 creator born April 1970-1973\\
\hspace*{1em}55. Shin Megami Tensei Digital Devil Saga development history first sketches 2002 2003 Wikipedia\\
\hspace*{1em}56. Shin Megami Tensei Digital Devil Saga Wikipedia full article development composer music\\
\hspace*{1em}57. Katsura Hashino Atlus director "born" OR "birthday" OR "birth date" April\\
\hspace*{1em}58. Persona 3 Wikipedia "first sketches" OR "initial concept" OR "concept art" development 2003\\
\hspace*{1em}59. 2007 in video games July releases list Wikipedia\\
\hspace*{1em}60. 2004 in video games July releases list Wikipedia\\
\hspace*{1em}61. Painkiller Xbox port release date 2006 exact date July\\
\hspace*{1em}62. Painkiller video game music composer born June 1968 1969 1970 1971\\
\hspace*{1em}63. Painkiller Hell Wars Xbox release date 2006 exact month\\
\hspace*{1em}64. Painkiller video game 2004 music composer soundtrack Wikipedia\\
\hspace*{1em}65. Painkiller video game 2004 music composer Marcin Przybyłowicz OR Tomasz Hajduk born June\\
\hspace*{1em}66. Painkiller video game People Can Fly development history first sketches 2002 2003 Adrian Chmielarz\\
\hspace*{1em}67. Painkiller video game Wikipedia full article development music sound\\
\hspace*{1em}68. Painkiller 2004 video game soundtrack composer Tomasz Hajduk OR Marcin Przybyłowicz OR Pawel Blaszczak\\
\hspace*{1em}69. Painkiller 2004 game music composer Tomasz Hajduk Wikipedia\\
\hspace*{1em}70. Adrian Chmielarz Painkiller "first sketches" OR "concept" OR "development began" 2002\\
\hspace*{1em}71. Painkiller video game 2004 Wikipedia "composer" OR "music" OR "soundtrack" Tomasz Hajduk\\
\hspace*{1em}72. Katsura Hashino Wikipedia "born" April 1970 1971 1972 1973 Atlus director\\[6pt]
\textbf{Perplexity final answer (incorrect):} fixates on Katsura Hashino / Persona~3 and Painkiller, concludes that Wikipedia "doesn't have Katsura Hashino's birth date," and never identifies \emph{LocoRoco} as the answer.
\end{tcolorbox}
\captionof{figure}{Full tool-call traces for one BrowseComp-Wikipedia query (gold page: \emph{LocoRoco}); the question is paraphrased to avoid benchmark leakage. SIRA grounds the indirect clues into category-anchored BM25 calls and retrieves the gold page at rank~1 in 11 retrieval calls (the winning call is marked $\star$). The agentic Perplexity-search baseline issues 72 distinct web searches, cycles through incorrect candidates (Persona~3, Painkiller), never surfaces the gold page, and commits to a wrong answer.}
\label{fig:browsecomp_case_study}

\end{document}